\documentstyle[prb,twocolumn,aps,tighten,floats,epsfig]{revtex}
\begin{document}
\draft 
\title{{\em Ab initio} Hartree-Fock computation of electronic 
static  structure factor of crystalline  insulators: benchmark
results on LiF} 
\author{Alok Shukla\cite{nadd}\cite{email}} 
\address{Max-Planck-Institut f\"ur
Physik komplexer Systeme,     N\"othnitzer Stra{\ss}e 38 
D-01187 Dresden, Germany}

\maketitle

\begin{abstract}
In this paper we present a fully {\em ab initio} Hartree-Fock 
approach aimed at calculating the static structure factor of 
crystalline insulators at arbitrary values of
momentum transfer. In particular, we outline the
computation of the incoherent scattering function, the component
of the structure factor which governs the incoherent x-ray scattering
from solids. The presented theory is applied to 
crystalline LiF to obtain benchmark Hartree-Fock values for its
incoherent scattering function. 
Benchmark theoretical values such as this, can be combined with the
experimentally measured static structure factor, to understand
the influence of electron correlation effects on cohesive properties
of solids. 
\end{abstract}
\pacs{ }
\section{INTRODUCTION}
\label{intro}
In order to obtain an {\em ab initio} understanding of the electronic
structure of solids, it is essential to understand the nature of
electron correlations in them~\cite{fulde}. Correlation being a real-space
phenomenon, one possible way of visualising  it is through the density-density
correlation function  defined as 
\begin{equation}
S({\bf r}', {\bf r}) = \frac{1}{N_0} \langle \Phi | \hat{\rho}({\bf r}') 
\hat{\rho}({\bf r})| \Phi \rangle \; \mbox{,}
\label{eq-dcorr}
\end{equation}
where $| \Phi \rangle$ denotes the many-particle wave function of the system, $N_0$ is the total number of electrons in the system and 
$\hat{\rho}({\bf r}) = \sum_{i=1}^{N_0} \delta ({\bf r}-{\bf r}_i)$ is
the density operator with ${\bf r}_i$ being the coordinates of the $i$-th electron. One can easily show 
\begin{equation}
S({\bf r}' , {\bf r}) = \delta ({\bf r}'-{\bf r}) + (N_0-1)
g({\bf r}' ,{\bf r}) \; \mbox{,}
\label{eq-dcorr2}
\end{equation}
where $g({\bf r}' , {\bf r})$
is the electron pair-correlation
function defined as
\begin{equation}
g({\bf r}',{\bf r}) = \frac{1}{\rho({\bf r})\rho({\bf r}')}
 \langle \Phi | \sum_{ i \neq j} \delta ({\bf r}'-{\bf r}_i) 
\delta ({\bf r}-{\bf r}_j) | \Phi \rangle \; \mbox{.}
\label{eq-pcorr}
\end{equation}
Above $\rho({\bf r})$ denotes
the electronic charge density. Pair-correlation function represents the
probability that when one electron is observed say at point ${\bf r}'$,
another electron will be found in a characteristic volume $1/(N_0-1)$
located at ${\bf r}$~\cite{fulde}. 
Clearly it can be used to 
quantify the so-called ``exchange-correlation hole" associated with an
electron in a many-electron system~\cite{fulde}.
If we compute the Fourier transform of the density-density correlator, we
obtain the static structure factor $S({\bf Q})$ defined by 
\begin{eqnarray}
S({\bf Q}) & = &\frac{1}{N_0} \int d{\bf r} d{\bf r}' e^{i{\bf Q}
\cdot ({\bf r} - {\bf r}')} S({\bf r}', {\bf r})
                \nonumber \\
          & = & \frac{1}{N_0} \langle \Phi | \sum_{j,k}e^{i{\bf Q}
\cdot ({\bf r}_{j} - {\bf r}_{k})} | \Phi \rangle \; \mbox{.}
\label{eq-sqdef}
\end{eqnarray}  
If we assume that the system under consideration is a crystalline system with
$N$ unit cells, each of which has $Z$ electrons in it so that $N_0=NZ$, we
can decompose $S({\bf Q})$ as a sum of a ``coherent" and an 
``incoherent" part
\begin{equation}
S({\bf Q}) = \frac{N}{Z} \delta_{{\bf Q},{\bf G}} 
|F({\bf Q})|^2 + S_{inc}({\bf Q}) 
\; \mbox{,}
\label{eq-sq2}
\end{equation}
where ${\bf G}$ is a vector of the reciprocal lattice and the form factor 
$F({\bf Q})$ defined as 
\begin{equation}
F({\bf Q})= \sum_{j=1}^{N_0} 
\langle\Phi|e^{i{\bf Q}\cdot {\bf r}_j}|\Phi\rangle
\; \mbox{,}
\label{eq-fq}
\end{equation} 
can be easily seen to be the Fourier transform of the charge density of
the system, while  
\begin{equation}
S_{inc}({\bf Q})= \frac{1}{NZ} ( \sum_{j,k=1}^{N_0} 
\langle\Phi|e^{i{\bf Q}\cdot ({\bf r}_j-{\bf r}_k)}|\Phi\rangle
 - |F(\bf Q)|^2)
\label{eq-sq}
\end{equation}
is referred to as the incoherent scattering function in the literature.
It is intuitively obvious that being the expectation value of 
a two-electron operator, 
$S_{inc}({\bf Q})$ will be sensitive to electron
correlations in the crystal, while $F(\bf Q)$, which is a one-electron
operator, should be relatively insensitive to such effects. It is
easy to verify that $S_{inc}({\bf Q})$ satisfies limiting conditions
\begin{equation}
\lim_{Q \rightarrow 0} S_{inc}({\bf Q}) = 0 \; \mbox{,}
\label{eq-qzero}
\end{equation} 
and
\begin{equation}
\lim_{Q \rightarrow \infty} S_{inc}({\bf Q}) = 1 \; \mbox{.}
\label{eq-qinf}
\end{equation}

One can perform the measurement of the static structure factor of a 
many-electron system in a variety of experiments such as electron scattering 
and x-ray scattering. For isolated atoms and molecules both 
electrons~\cite{barb} and
x-rays~\cite{watan} are frequently used for such measurements, 
however, for crystalline
sytems, x-ray scattering appears to be the method of 
choice~\cite{schulke,sacch1,sacch2,sacch3,mazzone}. In such 
measurements, the quantity
${\bf Q}$ in equations above is identifed with the momentum transferred
by the incident particle (electron or photon), to the
many-electron system under investigation.  Keeping in mind the relationship
between the static structure factor and the pair-correlation function
(cf. Eqs. (\ref{eq-dcorr2}) and (\ref{eq-sqdef}) ), it
is thus possible to  obtain the pair-correlation function from these 
measurements. However, in what follows, we will devote exclusively on the
x-ray scattering based experiments. The coherent x-ray scattering 
(Bragg scattering), i.e.  
when the momentum transfer ${\bf Q}$ is equal to one of the vectors 
${\bf G}$ of the reciprocal 
lattice, is governed predominantly by the form factor $F(\bf Q)$. However, 
by concentrating on the measurements corresponding to
those values of momentum transfer which are not equal to 
any reciprocal lattice vector, one 
can---according to Eq.(\ref{eq-sq2})---directly 
measure the incoherent scattering function. The experiments which concentrate
on this region of ${\bf Q}$ correspond to incoherent x-ray scattering. For the
case of incoherent scattering of x-rays from a crystalline solid at finite
temperatures,  assuming that the energy of the 
incoming x-rays is much higher compared to the binding energies of the 
constituent electrons, but still
low compared to the rest energy of the electron $m_0 c^2$ (so that the
relativistic effects can be neglected), the scattering cross section,
for a solid composed of light elements,
can be approximated as~\cite{sacch1,sacch2,sacch3}
\begin{equation}
\frac{d\sigma}{d\Omega} \simeq r_{0}^2 N \left[ \left(\frac{k}{k_0}\right)^2 (\tilde{e_0}
\cdot \tilde{e})^2 S_{inc}({\bf Q}) + S_{TDS}({\bf Q}) \right] \; \mbox{,}
\label{eq-cross}
\end{equation} 
where $r_0$ is the classical electron radius, ${\bf k}_0$ and ${\bf k}$
are the wave vectors of incoming and outgoing photons, $\tilde{e_0}$ and
$\tilde{e}$ are the corresponding polarization vectors, ${\bf Q} = {\bf k} -
{\bf k}_0$ is the momentum transfer while 
$S_{TDS}({\bf Q})$ is the structure
factor  due to the thermal diffuse scattering (TDS) caused both by the
thermal, and the zero-point vibrations of the lattice.
Thus $S_{TDS}({\bf Q})$ quantifies the contribution of  phonons
to the x-ray scattering, and can be computed by taking one-phonon,
and higher order terms into account~\cite{sacch2,sacch3}.  
Therefore, by measuring the incoherent
x-ray scattering cross section for different values of
the momentum transfer ${\bf Q}$, combinded with the knowledge of
$S_{TDS}({\bf Q})$, one can, using Eq. (\ref{eq-cross}), extract
the incoherent scattering scattering function $S_{inc}({\bf Q})$ of 
the system under consideration. 

Sacchetti and coworkers~\cite{sacch1,mazzone}
have been the proponents of using the incoherent
x-ray scattering to measure the static structure factor of crystalline 
compounds, for its subsequent use in the analysis of electron correlation
effects. They have  
performed a series of accurate measurements of the static sturcture
factor of the metallic system Be~\cite{sacch2} and the
covalent system diamond~\cite{sacch3} to obtain their 
pair-correlation functions, and analyzed various contributions
to the ground state energies of these compounds. In their latest experiment
performed on crystalline LiF, they have, for the first
time, subjected an ionic system to a similar analysis~\cite{sacch-lif}. 
However, in order to quantify the contribution of electron correlation
effects to the experimentally measured static structure factor in
such experiments, benchmark Hartree-Fock (HF) results for the quantity
are needed. It is the purpose of this  paper to present a formalism
using which one can perform such benchmark HF calculations within
an {\em ab initio} framework. Indeed,  
Calzuola et al.~\cite{sacch-lif}, by comparing their experimentally measured
values of the static structure factor, to the benchmark HF values 
presented here, have estimated the correlation contribution to the 
cohesive energy  of LiF. Therefore, the aim of these calculations
is {\em not} to explain the experimental data, but rather to provide
a theoretical reference, with respect to which the  correlation 
effects can be quantified  in the experimentally measured quantities. 
The formalism for computing the static structure factor presented here
is based on a Wannier-function-based {\em ab initio} 
HF approach developed recently by us~\cite{shukla1,shukla2}.
The approach has since been applied to compute the ground state properties
of a number of ionic~\cite{albrecht,shukla3} and
covalent compounds~\cite{shukla4,shukla5}, including  the form 
factor ($F({\bf Q})$) of LiF~\cite{shukla2}.  

Remainder of the paper is organized as follows.  In section \ref{theory} 
we describe our formalism for the {\em ab initio} evaluation
of the incoherent scattering function within an 
HF  approach. An explicit
formula is presented which represents $S_{inc}({\bf Q})$ in terms of the 
Wannier functions of an infinite crystal. Our numerical results for LiF are
presented in section \ref{results} which are compared to the experimental
results of Calzuola et al.~\cite{sacch-lif} for the same compound. Finally our
conclusions are presented in section \ref{conclusion}.
\section{THEORY}
\label{theory}
Here we outline the evaluation of the incoherent scattering function
$S_{inc}({\bf Q})$ for an infinite crystalline insulator within an 
{\em ab initio} restricted Hartree-Fock (RHF) approach. Although, we are not 
aware of such a prior calculation for an infinite solid, we note that 
{\em ab initio} calculations are performed on a routine basis on 
isolated atoms~\cite{isf-atm1,isf-atm2,isf-atm3,isf-atm4,isf-atm5,isf-atm6,%
isf-atm7,isf-atm8,isf-atm9,isf-atm10,isf-atm11,isf-atm12,isf-atm13} 
and molecules~\cite{isf-mol1,isf-mol2,isf-mol3,isf-mol4,isf-mol5,isf-mol6}, 
both at the HF, and the correlated level. 
However, for condensed-matter systems, perhaps because of practical
difficulties associated with providing a wave-function-based {\em ab initio}
description of an infinite system, such calculations are either
performed assuming a jellium model for the electrons
of the system~\cite{schulke,sacch2},
or within the framework of the density functional theory which 
often involves phenomenological approximations~\cite{isf-dft}. 

We assume that the compound under consideration is a closed-shell crystalline
system whose RHF ground state can be described by $n_c$ doubly occupied
Wannier functions per unit cell, so that $Z=2n_c$.
If we use Greek indices $\alpha$, $\beta$ etc. to denote the Wannier 
functions localized in a given unit cell, the RHF wave function of the 
infinite crystal can be described as a Slater determinant composed of 
the infinitely many Wannier functions 
$\{ |\alpha({\bf R}_{j})\rangle; \alpha =1,n_{c}; j=1,N \}$,
where $|\alpha({\bf R}_{j})\rangle$ denotes the 
$\alpha$-th Wannier function of a unit cell located at the position given 
by the vector ${\bf R}_{j}$ of the lattice. The aforementioned Wannier 
functions are assumed to form an orthonormal set
\begin{equation}
\langle\alpha({\bf R}_i ) | \beta ({\bf R}_j) \rangle = \delta_{\alpha\beta} \delta_{ij}
\mbox{,}
\label{ortho}
\end{equation}
and Wannier functions localized in different unit cells are translated
copies of each other
\begin{equation}
|\alpha({\bf R}_{i}+{\bf R}_{j})\rangle = {\cal T} ({\bf R}_{i}) 
|\alpha({\bf R}_{j})\rangle \mbox{,}
\label{eq-trsym}
\end{equation}
where  the operator ${\cal T} ({\bf R}_{i})$ 
represents a translation by lattice vector ${\bf R}_{i}$. The theory and
several applications of our approach, which directly obtains
the RHF Wannier functions of a crystalline insulator (ionic or covalent),
have been presented in several papers~\cite{shukla1,shukla2,albrecht,%
shukla3,shukla4,shukla5}.

If we use the standard formula for the expectation value of a two-particle
operator with respect to a Slater determinant~\cite{lindgren}, one can,  
after some algebraic manipulations, show that in the Wannier representation 
the RHF expression
for the incoherent scattering function (cf. Eq.(\ref{eq-sq})) is given by
\begin{equation}
S_{inc}({\bf Q}) = 1 - \frac{2}{Z} \sum_{\alpha,\beta=1}^{n_c} \sum_{i=1}^{N}
        |\langle \beta({\bf R}_{i}) | e^{i{\bf Q}\cdot {\bf r}} | \alpha({\bf
        0}) \rangle | ^2 \; \mbox{,}
\label{sq-fin}
\end{equation}
where  $|\alpha({\bf 0}) \rangle$ represents a Wannier function localized in
the reference unit cell. Eq. (\ref{sq-fin}) constitutes the key formula
of this work, a detailed derivation of which is presented in the appendix.
Since the Wannier functions in our  computer code
are represented in terms of Gaussian lobe-type localized basis functions,
it is possible to write down analytic expressions for the matrix elements
needed to evaluate $S_{inc}({\bf Q})$ according
to the expression above. The lattice sum over lattice vectors ${\bf R}_{i}$
involved in Eq. (\ref{sq-fin}) decreases rapidly as one moves away from the
reference cell, and is terminated once the convergence within a given threshold
is achieved.

The restriction of the present approach to 
insulators stems from our use of Wannier functions as the single-particle
orbitals, rather than the conventional Bloch orbitals. It is easy to see 
that the same theory can be easily extended to metallic systems if one were
to express the many-body wave function of the solid in terms of Bloch
orbitals. In that case, of course, the real-space sum (cf. Eq.(\ref{sq-fin}))
will have to be replaced by an integration over the Brillouin zone. We will
present this generalization in a future paper. This will be particularly
useful in light of the future experiments which Sacchetti and coworkers
are planning on metallic systems~\cite{sacch-met}.
\section{RESULTS AND DISCUSSION}
\label{results}
In this section we present the results of our HF calculations of the
incoherent scattering function at different values of the momentum transfer,
and compare our results to the experimental ones. It is intuitively
obvious, however, that HF structure factors can only describe the
experimental results qualitatively---in order to obtain a better
quantitative description theoretically, inclusion of 
electron correlation effects is essential. Nevertheless, in our opinion,
the comparison with experiments is very instructive, because one can,
in a rather pictorial way, see the successes and failures of the HF 
approximation in describing the physics of weakly-correlated systems.

The basis set used
to represent the Wannier functions in our calculations was the lobe 
representation of the basis set proposed by Prencipe et al. 
in their Bloch-orbital-based HF study of the structural properties of 
LiF~\cite{prencipe}. The basis set consisted  of contracted Cartesian 
Gaussian-type 
basis functions and was of [4s,3p] type for the fluorine atom, and
[2s,1p] for the lithium atom. For further details pertaining to the exponents
and the contraction coefficients we refer to the original work~\cite{prencipe}.
Details dealing with the lobe representation of the Cartesian Gaussian basis 
functions can be found, e.g, in our previous paper~\cite{shukla2}. We also
examined the basis-set dependence of our results on $S_{inc}({\bf Q})$ by
performing calculations with larger basis sets which also included d-type
basis functions on F atom, however, we did not observe any significant change
in the results. Thus we believe that our results on $S_{inc}({\bf Q})$
presented below are fairly accurate.
 
In the theoretical calculations the observed  
face-centered cubic (fcc) structure was assumed for the compound. 
The reference unit cell was taken to be the primitive cell
with the F atom at $(0,0,0)$ position and the Li atom at 
$(0,0,a/2)$, where $a$ is the lattice constant. For the lattice constant,
the room temperature value of $ 4.02 \AA$ was used. 

\begin{figure}
\psfig{file=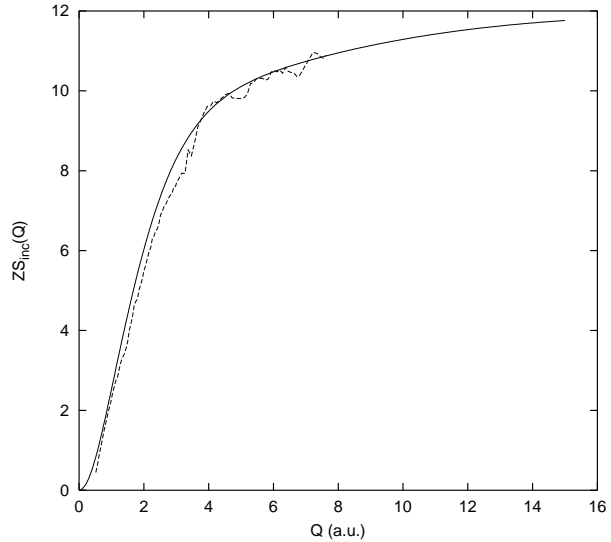,width=8.0cm,angle=0}
\caption{ $ZS_{inc}(Q)$ plotted as a function of the momentum transfer
$Q$ in the direction [100]. Solid line represents the 
HF theoretical results of this work, while
the dashed line represents the experimental results of 
Calzuola et al.~\protect\cite{sacch-lif}.  The experimental data was confined
to the values of $Q$  ranging from 0.517 a.u. to 7.622 a.u.}
\label{fig-sq} 
\end{figure}
\begin{figure}
\psfig{file=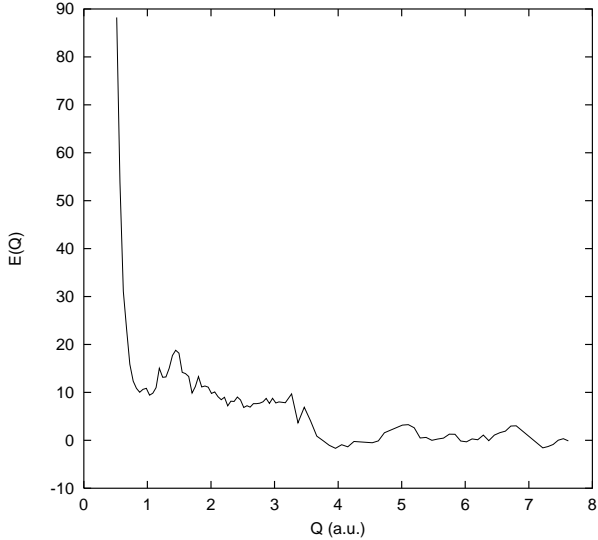,width=8.0cm,angle=0}
\caption{Relative correlation contribution $E(Q)$ to the measured
$S_{inc}({\bf Q})$, plotted as a function of the momentum
transfer $Q$. The momentum transfer direction was [100]. See text for details.}
\label{fig-eq} 
\end{figure}

The results of our calculations are
  are summarized in table \ref{tab-sq}
and Fig. \ref{fig-sq} which present $ZS_{inc}({\bf Q})$ as a function
of the momentum transfer ${\bf Q}$, where $Z=12$ for the case of LiF. 
Direction of the momentum transfer for both the
theory and the experiment was along [100] direction. Theoreitcal HF values of 
$S_{inc}({\bf Q})$ were calculated  with the magnitude of the
momentum transfer $Q$  ranging from 0.0 to 15.0 atomic units (a.u.). For the
experimental data,  $Q$ ranged from 0.517 a.u. to 7.622 a.u. It is quite
clear from Fig. \ref{fig-sq} that HF theory is in good qualitative agreement
with the experimental result, which is a manifestation of the fact that
LiF is a weakly correlated system.
\begin{table}  
 \protect\caption{Comparison of the Hartree-Fock incoherent scattering
function ($ZS_{inc}(Q)$) computed in this work with those measured
in the experiment of Calzuola et al.~\protect\cite{sacch-lif} at selected
values of momentum transfer ${\bf Q}$. The momentum transfer was
along the [100] direction in both the experiment and the theory.  }
  \begin{tabular}{ccc} \hline
 \multicolumn{1}{c}{Q} & \multicolumn{2}{c}{$ZS_{inc}(Q)$} \\ 
 (a.u.)   & This Work & Experiment  \\
\hline
  0.000     & 0.0000   &  --- \\
  0.100     & 0.0337   &  --- \\
  0.200     & 0.1352   &  --- \\
  0.300     & 0.2993   &  --- \\
  0.400     & 0.5188   &  --- \\ 
  0.517   & 0.8346   &  0.4434  \\
  0.569   & 0.9919   &  0.6446  \\ 
  0.621   & 1.1579   &  0.8832 \\
  0.672   & 1.3279   &  1.0774 \\
  0.724   & 1.5075   &  1.3008 \\ 
  1.034   & 2.6569   &  2.4282 \\ 
  1.137   & 3.0499   &  2.7480 \\
  1.291   & 3.6321   &  3.2082 \\ 
  1.497   & 4.3856   &  3.7112 \\    
  1.650   & 4.9182   &  4.3404 \\     
  1.804   & 5.4265   &  4.7896 \\
  1.957   & 5.9019   &  5.3122 \\     
  2.110   & 6.3461   &  5.8166 \\      
  2.263   & 6.7583   &  6.3042 \\     
  2.465   & 7.2525   &  6.6888 \\      
  2.768   & 7.8896   &  7.3224 \\     
  3.169   & 8.5596   &  7.9372 \\
  3.368   & 8.8304   &  8.5186 \\
  3.467   & 8.9521   &  8.3750 \\
  3.763   & 9.2716   &  9.2784 \\
  3.958   & 9.4506   &  9.6102 \\
  7.219   & 10.7878  & 10.9616 \\ 
  7.301   & 10.8060  & 10.9484 \\
  7.382   & 10.8237  & 10.9196 \\ 
  7.622   & 10.8747  & 10.8880 \\ 
  9.000   & 11.1331  &  ---  \\
 10.000   & 11.2894  &  --- \\
 12.000   & 11.5336  & --- \\
 14.000   & 11.7009  & --- \\
 15.000   & 11.7618  & --- \\
   \end{tabular}                      
  \label{tab-sq}    
\end{table}  
In order to quantify the correlation
effects,  we define the quantity 
$E({\bf Q}) = \frac{ S^{HF}_{inc}({\bf Q})-S^{exp}_{inc}({\bf Q})}
{S^{exp}_{inc}({\bf Q})} \times 100 $, which clearly measures the
percentile contribution of electron correlation effects to the experimentally
measured $S_{inc}({\bf Q})$, using the HF values presented here as
the benchmark reference.
 $E({\bf Q})$ is plotted in 
Fig. \ref{fig-eq}, as a function of $Q$.
For the smallest value of the momentum transferred measured $Q=0.517$ a.u.,
the correlation contribution is $88.2 \%$. With the increasing momentum
  transfer the correlation contribution decreases
rapidly staying in the range $10.0$ \% --- $20.0$ \% from  $Q=0.724$ a.u.
to $Q=2.059$ a.u.. From $Q=3.664$ a.u. onwards the uppper bound for 
the correlation contribution 
is approximately three percent, while most of the points are in one to 
two percent range. We also see some oscillations in the $E({\bf Q})$
as a function of the momentum transfer, which may be due to the experimental
uncertainties. However, the general trend in $E({\bf Q})$
as a function of the momentum transfer ${\bf Q}$ is clear---the contribution
due to the correlation effects decreases
with the increasing momentum transfer. This trend is also observed in
the calculations involving free atoms and molecules where the HF calculations 
for small values of momentum transfer always overestimate  $S_{inc}({\bf Q})$
as compared to the correlated ones~\cite{isf-atm1,isf-atm2,isf-atm3,%
isf-atm4,isf-atm5,isf-atm6,isf-atm7,isf-atm8,isf-atm9,isf-atm10,isf-atm11,%
isf-atm12,isf-atm13,isf-mol1,isf-mol2,isf-mol3,isf-mol4,isf-mol5,isf-mol6}.
This trend can be understood as follows. HF theory, because of a lack
of correlations in it, will always overestimate  the pair-correlation function
$g({\bf r}', {\bf r})$ and consequently $S_{inc}({\bf Q})$. 
Since for small values of momentum transfer $Q$, one can only probe the valence
electrons, the 
main contribution to $S_{inc}({\bf Q})$ will also naturally come from these  
electrons. However, it is the valence electrons for which the 
correlation effects are quantitatively the most important, and their
neglect in the HF approach leads to relatively large deviations, as compared
to the experimental values, for smaller values of momentum transfer. 

In our calculations we did not observe any significant anisotropy in 
$S_{inc}({\bf Q})$ with respect to the direction of ${\bf Q}$. We performed
the same set of calculations for $S_{inc}({\bf Q})$ for momentum transfer
directions [110] and [111] as well, however, the difference in the
results compared to [100] direction was always less than 
$1.0 \times 10^{-5}$. This result can also be understood on intuitive grounds
as the charge density in LiF is fairly isotropic, therefore, one would not
expect the incoherent scattering function to show any anisotropy.
Finally it is clear from both table \ref{tab-sq} and Fig. \ref{fig-sq} that
our HF results on $S_{inc}({\bf Q})$ approach the correct limiting values
in both the low momentum-transfer region (cf. Eq.(\ref{eq-qzero})) and
the high momentum-transfer region (cf. Eq.(\ref{eq-qinf})). 
\section{CONCLUSIONS}
\label{conclusion}
In conclusion a wave-function-based fully {\em ab initio} approach has been 
presented, using which, one can compute the static structure factor of a 
crystalline compound at arbitrary values of the momentum transfer, at the
HF level. The formalism was applied to the case of crystalline LiF, and
benchmark values were obtained for its incoherent scattering function.
These values were used in the analyssis of a recently performed 
incoherent x-ray scattering experiment on LiF to quantify the electron 
correlation effects, in general,
and to predict the correlation contribution to its cohesive energy,
in particular~\cite{sacch-lif}. In case such experimental measurements 
are performed on other insulating compounds,
one can use the formalism presented here to perform similar benchmark 
calculations on those systems as well. The present version of the theory
is restricted to insulating systems because of its Wannier-function-based
formulation, however, in light of the planned future experiments on metallic
systems~\cite{sacch-met}, we do intend to develop an {\em ab initio} HF 
formalism meant for computing $S_{inc}({\bf Q})$ for gapless
systems, as well.

Although the aim of the present HF formalism was not to explain the 
experimental data,
but rather to facilitate its analysis, it is still of interest to
include electron correlation effects theoretically. 
Indeed, recently, we have generalized our Wannier-function-based approach
to include electron correlation effects by systematically enlarging
the many-particle ground-state wave function by considering virtual
excitations from the space of the occupied Wannier functions to that
of the virtual ones~\cite{shukla-corr}. The approach was demonstrated by
computing the correlation contributions to the total energy per unit
cell of bulk LiH~\cite{shukla-corr}. However, the generalization of the
approach to compute the correlated expectation value of an operator
other than the Hamiltonian is far from trivial, and will be the subject
of a future investigation.

\acknowledgements
It is my pleasure to thank Professor F. Sacchetti for bringing his experiments 
to my attention, and for encouraging me to do these calculations. 
I would also like to thank Professor P. Ziesche for his thorough reading
 of the manuscript, and for providing an exhaustive
list of references on the subject. Thanks are also due to Sumit
Mazumdar for his critical comments.
Finally, I would like to express my gratitude to Professor P. Fulde for 
supporting my research
throughout my stay in the Max-Planck-Institut f\"ur Physik komplexer 
Systeme, Dresden.
\appendix
\section*{}
Our aim in the present section is to present a derivation of Eq.(\ref{sq-fin})
of the text which is an RHF level expression for
$S_{inc}({\bf Q})$ in terms of corresponding Wannier functions. 
Eq.(\ref{eq-sq}) which defines $S_{inc}({\bf Q})$ involves the expection
value of the operator
\begin{equation}
X({\bf r}_1,{\bf r}_2,\ldots,{\bf r}_{N_0};{\bf Q}) 
= \sum_{j,k=1}^{N_0} 
e^{i{\bf Q}\cdot ({\bf r}_j-{\bf r}_k)} \: \mbox{.}
\label{eq-xop}
\end{equation}
Although operator $X({\bf r}_1,{\bf r}_2,\ldots,{\bf r}_{N_0};{\bf Q})$ is
a sum of two-electron terms, however, unlike other similar operators
such as the Coulomb interaction operator, the sum in Eq.(\ref{eq-xop})
does contain the term where $j=k$. Therefore, in order to utilize the
well-established formulas for matrix elements of two-electron operators
between Slater determinant~\cite{lindgren}, we rewrite Eq.(\ref{eq-xop})
as 
\begin{eqnarray}
X({\bf r}_1,{\bf r}_2,\ldots,{\bf r}_{N_0};{\bf Q}) 
 & = & N_0 + Y({\bf r}_1,{\bf r}_2,\ldots,{\bf r}_{N_0};{\bf Q}) \nonumber \\
&   & +Y^{*}({\bf r}_1,{\bf r}_2,\ldots,{\bf r}_{N_0};{\bf Q}) \; \mbox{,}
\label{eq-xop2}
\end{eqnarray}
where the first term on the right-hand side corresponds to $j=k$ terms of
the sum in Eq.(\ref{eq-xop}) and the operator $Y$ is defined as
\begin{equation}
Y({\bf r}_1,{\bf r}_2,\ldots,{\bf r}_{N_0};{\bf Q})  =  
\sum_{j<k} e^{i{\bf Q}\cdot ({\bf r}_j-{\bf r}_k)} \mbox{.}
\label{eq-yop}
\end{equation}
$Y^{*}$, which represents the complex conjugate of operator $Y$, can be
easily deduced from Eq.(\ref{eq-yop}). It is clear that $Y$ 
(and hence, $Y^{*}$) as defined above are traditional two-electron operators.
Utilizing the well-known formula for the expectation value of a 
general two-electron
operator $G=\sum_{j<k} g({\bf r}_j,{\bf r}_k)$ with respect to a single 
Slater determinant state $|\Psi\rangle$~\cite{lindgren}
\begin{eqnarray}
\langle\Psi|G|\Psi\rangle & = & \frac{1}{2} \sum_{a,b} \langle a b|
g(({\bf r}_1,{\bf r}_2)|a b \rangle  \nonumber \\
&   & - \frac{1}{2} \sum_{a,b} \langle b a|
g(({\bf r}_1,{\bf r}_2)|a b \rangle \: \mbox{,}
\label{eq-gop}
\end{eqnarray}
where $| a \rangle$, $| b \rangle$ are the orbitals constituting the Slater
determinant, one can easily get
\begin{eqnarray}
\langle\Phi|Y|\Phi\rangle & = & \sum_{\alpha, \beta =1}^{n_c} 
\sum_{{\bf R}_{i},{\bf R}_{j} =1}^{N}  \left\{
2 \langle \alpha({\bf R}_{i})| e^{i{\bf Q}\cdot {\bf r}}|\alpha({\bf R}_{i})
\rangle \right. \nonumber \\
&   & \left. \times \langle \beta({\bf R}_{j})| e^{-i{\bf Q}\cdot {\bf r}}|\beta({\bf R}_{j})
\rangle -  \langle \beta({\bf R}_{j})| e^{i{\bf Q}\cdot {\bf r}}
|\alpha({\bf R}_{i}) \rangle \right. \nonumber \\ 
&   & \left. \times \langle \alpha({\bf R}_{i})| 
e^{-i{\bf Q}\cdot {\bf r}}|\beta({\bf R}_{j})
\right\} \: \mbox{,} \label{eq-yop2}
\end{eqnarray}
where $|\Phi\rangle$ represents the single-Slater-determinant RHF ground
state of the crystalline insulator under consideration. As discussed in
Sec. \ref{theory}, it is expressed in terms of Wannier functions 
$\{ |\alpha({\bf R}_{j})\rangle; \alpha =1,n_{c}; j=1,N \}$, where $2n_{c}$
is the number of electrons per unit cell and $N$ is the total number unit
cells considered. In order to arrive at the right-hand side of 
Eq.(\ref{eq-yop2}) spin summations have been performed, and the first term
represents the so-called ``direct" contribution, while the second term 
represents the ``exchange'' contribution. It is clear that both the direct,
 as well as the exchange terms are expressed as products of two one-electron
matrix elements which are complex conjugates of each other. Moreover, using 
the defining Eq.(\ref{eq-fq}),
one can easily see that, for the RHF state considered here, the direct
term of Eq.(\ref{eq-yop2}) is nothing but the product of the form factor 
$F({\bf Q})$ and its complex conjugate, leading to 
\begin{eqnarray}
\langle\Phi|Y|\Phi\rangle & = &  \frac{1}{2} |F({\bf Q})|^2 \nonumber \\
&    & - 
\sum_{\alpha, \beta =1}^{n_c} \sum_{{\bf R}_{i},{\bf R}_{j} =1}^{N} 
 |\langle \beta({\bf R}_{j})| e^{i{\bf Q}\cdot {\bf r}}
|\alpha({\bf R}_{i}) \rangle |^2
 \: \mbox{.} \label{eq-yop3}
\end{eqnarray}
Finally, using the translational invariance property of the 
one-electron matrix elements $\langle \beta({\bf R}_{j})| e^{i{\bf Q}\cdot 
{\bf r}}|\alpha({\bf R}_{i}) \rangle = \langle \beta({\bf R}_{j}-{\bf R}_{i})
| e^{i{\bf Q}\cdot {\bf r}}|\alpha({\bf 0}) \rangle$, and by rearranging
the sum over lattice vectors ${\bf R}_{i}$ and ${\bf R}_{j}$ in 
Eq.(\ref{eq-yop3}), one gets in the infinite solid limit 
($N \rightarrow \infty$)  
\begin{equation}
\langle\Phi|Y|\Phi\rangle =  \frac{1}{2} |F({\bf Q})|^2 - N 
\sum_{\alpha, \beta =1}^{n_c} \sum_{{\bf R}_{i} = 1}^{N} 
 |\langle \beta({\bf R}_{i})| e^{i{\bf Q}\cdot {\bf r}}
|\alpha({\bf 0}) \rangle |^2
 \: \mbox{.} \label{eq-yopfin}
\end{equation}
If we combine the results of Eqs. (\ref{eq-xop}),(\ref{eq-xop2}), 
(\ref{eq-yopfin}), and substitute them in Eq.(\ref{eq-sq}), we immediately
obtain Eq.(\ref{sq-fin}), valid for a single Slater determinant
RHF wave function $|\Phi\rangle$.

%
%
%
%
%
\end{document}